\documentstyle[aps, floats, prl]{revtex}

\newcommand{\bastar}{\begin{eqnarray*}}
\newcommand{\eastar}{\end{eqnarray*}}
\newskip\humongous \humongous=0pt plus 1000pt minus 1000pt

\newif\ifdtup

\relax
\newcommand{\be}{\begin{equation}}
\newcommand{\ee}{\end{equation}}
\newcommand{\bea}{\begin{eqnarray}}
\newcommand{\eea}{\end{eqnarray}}
\newcommand{\X}{{\bar X}}
\newcommand{\pro}{\partial}
\newcommand{\n}{\hat n}
\newcommand{\oneg}{\displaystyle\frac{1}{g}}

\newcommand{\tone}{{{\theta}_1}}
\newcommand{\ttwo}{{{\theta}_2}}

\newcommand{\C}{{\vec C}}
\newcommand{\vC}{{\vec C}}
\newcommand{\vth}{{\vec \theta}}
\newcommand{\vX}{{\vec X}}

\newcommand{\Xb}{{\bar X}}

\newcommand{\D}{{\hat D}}
\newcommand{\Db}{{\bar D}}

\newcommand{\hA}{\hat A}

\newcommand{\A}{{\vec A}}

\newcommand{\dfrac}{\displaystyle\frac}
\newcommand{\ba}{\begin{array}}
\newcommand{\ea}{\end{array}}

\newcommand{\nn}{\nonumber}
\begin{document}
\twocolumn[\hsize\textwidth\columnwidth\hsize\csname@twocolumnfalse%
\endcsname
\title  { Effective Lagrangian with valence gluons in 
extended $SU(2)$ QCD model}
\bigskip
\author{Y. M. Cho$^{1,2}$, Haewon Lee$^{3}$, D. G. Pak$^{1,4}$}
\address{
$^{1)}$Asia Pacific Center for Theoretical Physics,
Seoul 130-012, Korea \\
$^{2)}$Department of Physics, College of Natural Sciences, Seoul National 
University,
Seoul 151-742, Korea
\\$^{3)}$Department of Physics, Chungbuk National University, Cheongju, Chungbuk 361-763, Korea
\\ $^{4)}$Department of Theoretical Physics, Uzbekistan National University,
Tashkent 700-174, Uzbekistan\\
{\scriptsize \bf ymcho@yongmin.snu.ac.kr, hwlee@phys.chungbuk.ac.kr,
 dmipak@mail.apctp.org} \\ \vskip 0.3cm
}

\maketitle 

\begin{abstract}

   The effective action with homogeneous valence 
(off-diagonal) gluons as background fields 
in the extended $SU(2)$ model of QCD is
obtained in one-loop approximation. We keep the manifest gauge 
and Lorentz invariance during whole calculation by using a special 
gauge-fixing recipe and taking  into account  
the quartic interaction terms in the original Lagrangian as well. 
It has been shown that the effective Lagrangian  
gains  a positive imaginary part.

\vspace{0.3cm}
PACS numbers: 12.38.-t, 11.15.-q, 12.38.Aw, 11.10.Lm
\end{abstract}

\narrowtext
\bigskip
                          ]

    The establishment of color confinement and 
derivation of the low energy  effective theory
 from the first principles is one of important   problems in quantum 
chromodynamics
(QCD). For last years the 
 idea of monopole
condensation which can  provide the color confinement mechanism \cite{nambu,thooft} has
been  supported in analytical studies \cite {ezawa} and lattice
simulations \cite {kronf}.
 Recently it was shown  \cite{q1} in the framework of
the   extended $SU(2)$ QCD model 
\cite{cho1} that the effective theory obtained after integrating out 
all dynamical degrees of freedom can describe the confinement phase with the monopole condensation and 
dynamical gauge symmetry breaking. The later  provides the effective mass generation and gives the origin for 
 the kinetic term in the Skyrme-Faddeev effective Lagrangian \cite{fadd1,niemi,q1}.
It was proposed \cite{q1} that the effective action 
has a new duality symmetry which is closely related with
the vacuum stability in  the case of a 
pure   chromomagnetic background whereas in the presence of a chromoelectric background 
the corresponding vacuum becomes unstable due to  the imaginary part of
the effective Lagrangian. The imaginary part for chromomagnetic and 
chromoelectric background in standard $SU(2)$ model of QCD was considered
in \cite{savv,niel,raji,schan}.

In the present paper we study further the quantum properties of the
 extended $SU(2)$ model of QCD  and 
calculate the one-loop
effective action with covariant constant valence 
(off-diagonal) gluons as background fields.
Notice, that the extended  $SU(2)$  model of QCD \cite{cho1} 
incorporates naturally the topological scalar field
which insures the off-diagonal gluon to be represented as a gauge covariant 
object in the extended gauge theory.
This allow us to treat the effective action with off-diagonal gluons  in 
a consistent gauge invariant manner.
Our effective Lagrangian
with covariant constant
valence gluons reveals a non-trivial vacuum but possesses a 
positive imaginary part.
We maintain the explicit Lorentz and gauge invariance during calculation
of the effective Lagrangian
and take into account the contribution which is coming from 
the quartic gluon interaction terms in the classical
 Lagrangian.

 We will follow the original papers  \cite{cho1} and remind
 first the main lines 
of the extended $SU(2)$ model of QCD.
     The  principal construction in that model
 is a generalized parametrization for the $SU(2)$ gauge connection 
implemented with a  scalar field $\n$ which describes
 the topological degrees of freedom of the non-Abelian gauge theory
\bea
  \vec{A}_\mu &=&A_\mu \n - \oneg \n\times\pro_\mu\n+\vX_\mu\nonumber
  \\          &=& \hat A_\mu + \vX_\mu, \,\,\,\,\,\,  (\n^2 =1).
\eea
The restricted conection $\hat A_\mu$ still represents an $SU(2)$ connection which
undergoes the  full $SU(2)$ group gauge 
transformations and at the same time maintains 
all topological characteristics of the original unconstrained  non-Abelian potential.
The isolated singularities of $\hat{n}$ define
the second homotopy group
 $\pi_2(S^2)$
which describes the non-Abelian monopoles from the different topological
sectors.  Indeed, $\hat A_\mu$
with $A_\mu =0$ and $\hat n= \hat r$ leads to 
 Wu-Yang point-like monopole solution \cite{wu,cho3}.  Besides that,
 with the $S^3$
compactification of the three-dimensional space $R^3$, the all mappings $\hat{n}$
are characterized by the
Hopf invariant $\pi_3(S^2)\simeq\pi_3(S^3)$ which classifies  the topologically distinct vacua
in the non-Abelian gauge theory
\cite{bpst,thooft}. Notice, that since the $SU(2)$ connection space is  an affine space
one can extend the restricted connection ${\hat A}_\mu$ by adding a general covariant vector
 $\vX_\mu$ (so-called ``valence gluons'' \cite{cho1}) which corresponds to
the  off-diagonal components of the gauge potential  
in the standard  $SU(2)$ QCD.

Under the infinithesimal $SU(2)$ gauge transformation
\bea
\delta \n = - \vec \theta \times \n  \,,\,\,\,\,
 \delta \A_\mu = \oneg  {\vec D}_\mu \vec \theta,
\eea
one has the following transformation rules for the Abelian component $A_\mu$ 
(``photon''),
the restricted connection 
 $\hA_\mu$ 
and valence gluons $\vX_\mu$
\bea
\delta A_\mu &=& \oneg \n \cdot \pro_\mu \vth,\,\,\,\,\,\,\,\,\,
           \delta \hat A_\mu = \oneg \D_\mu \vth  ,  \nn \\
\delta \vX_\mu &=& - \vth \times \vX_\mu  ,
\eea
  here  
\bea
\D_\mu  = \pro_\mu  + g {\hat A}_\mu 
\eea
is a covariant derivative of the restricted theory.
It should be noted that the restricted connection
$\hA_\mu$ itself keeps the whole $SU(2)$ gauge invariance providing
by this the gauge independent structure of the
 Abelian projected effective  theory.
The Lagrangian of the extended Yang-Mills theory can be written as follows 
\cite{cho1}
\bea
{\cal L}_0& = &-\dfrac{1}{4} \vec F^2_{\mu \nu }
     =-\dfrac{1}{4}
{\hat F}_{\mu\nu}^2 -\dfrac{1}{2} {\hat F}_{\mu\nu} \cdot \n X_{\mu \nu}  \nn \\
 &&-\dfrac{1}{4} X_{\mu \nu}^2  -  \dfrac{1}{4} ( \D_\mu \vX_\nu - \D_\nu \vX_\mu)^2 ,
\eea
where the main field strengths for the ``photon'' $A_\mu$ , magnetic potential $\vC_\mu$ and 
valence gluons $\vX_\mu$ are defined by
\bea
{\vec F}_{\mu \nu} &=& \hat F_{\mu \nu} \n + \D_\mu \vX_\nu - \D_\nu \vX_\mu +
     g \vX_\mu \times \vX_\nu , \nn \\
\hat F_{\mu\nu} &=&  (F_{\mu\nu}+H_{\mu\nu})\n, \nn \\
F_{\mu\nu}&=&\pro_\mu A_\nu-\pro_\nu A_\mu,\,\,\, \nn  \\
X_{\mu \nu} &=& g \n \cdot (\vX_\mu \times \vX_\nu) , \nn \\
H_{\mu\nu}&=&-\oneg\n\cdot(\pro_\mu\n\times\pro_\nu\n ) \nn \\   
    &=&( \pro_\mu \C_\nu - \pro_\nu \C_\mu + g \C_\mu \times \C_\nu)\cdot \n .  
\eea
 The vector connection  $\C_\mu \equiv -  \oneg \n\times\pro_\mu\n$ is exactly the magnetic potential
which corresponds to the classical Wu-Yang monopole solution \cite{wu} or to Dirac string 
after making
gauge transformation to a singular magnetic gauge \cite{cho1}.
In the magnetic gauge the 
magnetic potential for the monopole solution is represented explicitly by an  Abelian part of the
full connection. To pass to the  magnetic gauge it is more convenient to decompose the 
 vector fields $\C_\mu $ and $\vX_\mu$ in the orthonormal basis $(\n^a, n^3 \equiv \n;
 a=1,2)$
instead of explicit rotating the internal space of $SU(2)$ as it was defined in \cite{cho1} 
\bea
\C_\mu = C_\mu^a \n^a, \,\,\,\,\,\,\,\, \vX_\mu = \vX_\mu^a \n^a.
\eea
After introducing complex 
notations for the vector fields $X_\mu = \dfrac{1}{\sqrt 2} (X_{1\mu} + i X_{2\mu})$,
        ${\bar X}_{\mu} = (X_\mu)^\ast $  one can rewrite the classical Lagrangian
${\cal L}_0$ as
\bea
 {\cal L}_0 &=& - \dfrac{1}{4} ({F}_{\mu \nu}+ H_{\mu \nu})^2  - \dfrac{1}{2}
  | D_\mu X_\nu - D_\nu X_\mu  |^2  \nn \\
 &&+ i g ({F}_{\mu\nu}+ H_{\mu \nu}) \X_\mu X_\nu - V(X^4), \nn \\
 V(X^4) &=&\dfrac{ g^2}{2} [  (\X X)^2 -  \X^2 X^2 ], \nn \\
 D_\mu &=& \pro_\mu + i g (A_\mu + C_\mu), \nn \\
C_\mu &=& -\oneg \n_1 \pro_\mu \n_2.   
\eea
The magnetic field strength $H_{\mu \nu}$ can be rewritten in the magnetic gauge 
as follows
\bea
H_{\mu\nu}=\pro_\mu C_\nu - \pro_\nu C_\mu. 
\eea

We have two types of gauge transformations  under which the original Lagrangian is invariant:\\
(I) the active type transformations defined by Eqs. (2) and (3)
 and (II) the passive ones which have the following form
\bea 
 \delta C_\mu &=& 0, \nn \\
 \delta A_\mu &=& \oneg \pro_\mu  \theta_3 + i {\bar \theta} X_\mu - i \theta \X_\mu, \nn \\
 \delta X_\mu &=& \oneg D_\mu \theta - i \theta^3 X_\mu .
\eea
It should be noted that
the passive type gauge transformations possess a structure of the complexified $U(1)$ gauge
group and  in addition 
 we have also the Abelian dual
$U_m (1)$ magnetic gauge symmetry for the magnetic potential $C_\mu$ \cite{cho1}.

 Let us    consider the generating 
functional for connected  Green functions with the sources corresponding to
the Abelian  and off-diagonal gluons
 \bea
W[J_\mu, {\vec J}_\mu]&=&
\int DA_\mu D \vX_\mu 
 \exp [- i \int  ( \dfrac{1}{4} {\vec F}_{\mu\nu}^2  \nn \\
&&+ A_\mu J_\mu +\vX_\mu \cdot {\vec J}_\mu) d^4 x  ] .
\eea
We split the gauge fields 
 into the classical background $A_{0\mu}, X_{0\mu}$ and
quantum parts $A'_\mu, X'_\mu$ 
\bea
A_\mu &=& {A}_{0\mu} + A'_\mu , \nn \\
X_\mu & = &{X}_{0\mu} + X'\mu. 
\eea
The magnetic potential $C_\mu$ (or equivalently the topological scalar field $\n$)
is treated as a classical object since it does not contain 
dynamical degrees of freedom in the original theory.
  Notice that formally we can include  the magnetic potential $C_\mu$
into the classical part $A_{0\mu}$ of  the Abelian
gauge component since the $C_\mu$ enters the covariant 
derivatives  $D_\mu, {\bar D}_\mu $ \, in additive manner
everywhere.
In computation of functional determinants in the non-Abelian gauge theory
while keeping the manifest Lorentz invariance 
one encounters an obstacle of calculating the matrix  determinants
with respect to Lorentzian and internal group indices. To overcome
this technical difficulty we choose the special gauge-fixing functions $f, h$ 
\bea 
 && f = D_{0 \mu} X'_\mu  - ig A'_\mu X_{0\mu} , \nn \\
&& h = \pro_\mu A'_\mu + i g X_{0\mu} \Xb'_\mu - i g \Xb_{0\mu} X'_\mu  ,
\eea
where $D_{0 \mu} = \pro_\mu + i g (A_{0\mu } + C_{\mu})$ is a $U(1)$ covariant derivative
with background  Abelian fields. 
This  choice allows us to factorize the functional determinants in Lorentzian indices.
Using the passive type gauge transformations and
applying the 't Hooft trick of inserting the unit into the generating functional 
 we obtain the gauge-fixing Lagrangian (in Feynman gauge) 
\bea 
{\cal L}_{gf} =  - f f^\ast - \dfrac{1}{2} h^2
\eea
and the Faddeev-Popov ghost determinant ${\rm Det} M_{FP}$ in a standard way.
 We calculate the effective action in one-loop approximation and 
 consider the   
background valence gluons which are covariant constant under the transformations
of the stability subgroup $U(1)$
\be
D_{0\mu} X_\nu = 0 .
\ee
 Since the background Abelian field $A_{0\mu} + C_\mu$
appears everywhere only in terms of the covariant background derivative
$D_{0\mu}$ and 
background field strength
$F_{0\mu\nu}$ we can rewrite the effective Lagrangian in an explicit  covariant form.
Our choice of the gauge-fixing functions, Eq. (13),   
 simplifies  crucially  the kinetic term structure
\bea 
{ L}_0 + { L}_{gf} &=& \dfrac{1}{2} A'_\mu [ g_{\mu\nu} ( \Box - 2 a ) ] A'_\nu \nn \\
&& +     2 i g A'_\mu (X_{0\nu}  \Db_{0\nu} \Xb'_\mu  
 - \Xb_{0\nu} D_{0\nu} X'_\mu) \nn \\
&& + \Xb'_\mu [ g_{\mu \nu} ( D_0 D_0 - a ) + 2 i g F_{0\mu\nu}] X'_\nu  \nn \\
&&+ 2g^2 \Xb'_\mu ( \Xb_{0\mu} X_{0\nu} 
-X_{0\mu } \Xb_{0\nu}) X'_\nu   \nn \\
 &&+ \dfrac{g^2}{2} \Xb'^2 X_0^2 + \dfrac{g^2}{2} X'^2 \Xb_0^2 ,
\eea
where $ a =g^2 \Xb_{0\mu} X_{0\mu}$ and we keep only terms
quadratic in quantum fields which are enough at one-loop
level. 
The Faddeev-Popov matrix operator is 
\bea
&&M_{FP} = \dfrac{\delta (f, \bar f, h)}{\delta (\theta, \bar \theta, \theta^3)}\nn \\
&& = \oneg \left( \begin{array}{ccc}
          D_0 D_0  - a                     & g^2  X_0^2                &  - 2 i g X_{0\mu} \pro_\mu    \\
     g^2  {\bar X}_0^2          & \Db_0 \Db_0 - a      & 2 i g \X_{0\mu} \pro_\mu     \\
         -2 ig \Xb_{0\mu} D_{0\mu}    & 2 ig X_{0\mu} \Db_{0\mu} & \Box -2 a          
\end{array} \right).
\eea
We consider the case of  $U(1)$ covariant constant valence gluon background.  
The integration over Abelian field $A_\mu $ results in a functional determinant
\bea 
 {\rm Det}^{-\frac{1}{2}} K_A =
 {\rm Det}^{-\frac{1}{2}} [g_{\mu\nu} (\Box - 2 a)]
\eea
and additional non-local terms which will contribute to the effective 
action after subsequent integrating out the
quantum parts $X'_\mu$ of the valence gluons.
 The final  expression for the functional determinant obtained
after performing the 
integration over Abelian and  off-diagonal gauge fields is the following
\bea
&& {\rm Det}^{-\frac{1}{2}} K_{\mu \nu} = {\rm Det}^{-\frac{1}{2}} 
\left ( \begin{array}{cc}
                   {\tilde A}   &  {\tilde B} \\
                    {\tilde C}   &  {\tilde D}
                    \end{array}  \right ) , \nn  \\
&&{\tilde A}_{\mu \nu} = g_{\mu\nu} (D_0 D_0 - a + 
\dfrac{\alpha}{\Box - 2 a} ) + 2 i g( F_{0\mu\nu} +X_{0\mu\nu} ), \nn \\
&&{\tilde B}_{\mu \nu} = g_{\mu \nu} (g^2 X_0^2 - \dfrac{\beta}{\Box - 2 a}) \equiv g_{\mu \nu } B, \nn \\
&&{\tilde C}_{\mu \nu}  = {({\tilde B}^*)}_{\mu \nu}\equiv g_{\mu \nu} C ,
                  \,\,\,\,\,\,\,  {\tilde D}_{\mu \nu}=
               { {\tilde  A}^*}_{\mu \nu}  .                      
\eea
where 
\bea
&&\alpha = 4g^2 \X_{0\mu} X_{0\nu} D_{0\mu} D_{0\nu} ,\nn \\
&&\beta = 4g^2 X_{0\mu} X_{0\nu} \Db_{0\mu} \Db_{0\nu} .
\eea
So that the one-loop contribution  to the effective action is given by
\bea
\Delta S_{eff} = -\dfrac{1}{2} \, {\rm Tr}\, \ln K_A 
      -\dfrac{1}{2} \, {\rm Tr} \, \ln K_{\mu \nu} + {\rm Tr} \, \ln M_{FP}.
\eea

For simplicity we
 consider the effective Lagrangian with off-diagonal background gauge fields
only, 
so from now on  we set  $A_{0\mu} = 0$.
The determinant of the matrix $K_{\mu \nu} $ can be  simplified 
to the next form
\bea 
{\rm Det} K_{\mu\nu} &=& {\rm Det} (\tilde A \tilde D - \tilde B \tilde C)  \nn \\
   &=& {\rm Det} [g_{\mu \nu} ( {A}^2 - BC) 
+  4 X_{0\mu\rho} X_{0\rho\nu}], 
\eea
where
\bea
{A} = \Box - a + \dfrac{\alpha}{\Box - 2 a} . 
\eea
The field strength $X_{\mu\nu}$ has a useful cyclic 
property
\bea 
X_{\mu\nu}X_{\nu\rho}X_{\rho\sigma} &=&-\dfrac{1}{2} { \chi}^2 X_{\mu\sigma}, \nn \\
{\chi} &=& \sqrt {X_{0\mu\nu} X_{0\mu\nu}}.
\eea
Using this relation one can calculate the part of the determinant in Eq. (22)
with respect to Lorentzian indices. After some simplifications 
one obtains 
\bea
&&\ln {\rm Det} K_{\mu\nu} \nn \\
&&= 2 {\rm tr} \, \ln ( { A}^2 - BC) + 2 {\rm tr} \, \ln ({ A}^2 - BC - 2 { \chi}^2 ) ,
\eea
where ``${\rm tr}$'' involves only the remaining functional trace.

The Faddeev-Popov determinant is simplified in a similar manner
\bea
{\rm Det} M_{FP} = {\rm Det} ( {A}^2 - BC) {\rm Det} ( \Box - 2 a) .
\eea
Summing up contributions from all determinants one obtains the  simple expression for the one-loop 
contribution to the effective action
\bea
\Delta S_{eff} = - {\rm tr} \, \ln ( {A}^2 - BC - 2 { \chi}^2) - {\rm tr} \, \ln (\Box - 2 a).
\eea
This  expression  has no neither Lorentzian nor internal group  indices and 
is quite suitable for  calculation of  the remaining functional traces.
Passing to the Eucledian space-time and making Fourier transformation
we can fullfil the  integration over momentum with using a cut-off
regularization scheme.
To obtain the final expression for the effective Lagrangian 
one needs to know only the functional dependence on the valence gluons. The easiest
way to find 
the functional form of the effective Lagrangian is to replace the complex vectors
$X_\mu, \X_\mu$ with a pair of  real vectors in Eucledian space-time 
\bea
& X_{\mu E} \rightarrow V_\mu , \,\,\,\,\,
\X_{\mu E} \rightarrow W_\mu .  
\eea
In four dimensional Eucledian  
spherical coordinates the momentum $k_\mu$ has the following projections 
\bea
    k_\mu =  
\left ( \begin{array}{ll}
                   k \cos \theta_1 \\
                   k \sin \theta_1 \cos \ttwo  \\
                   k  \sin \tone \sin \ttwo \cos \phi  \\
                   k\sin \tone \sin \ttwo \sin \phi
                    \end{array}  \right ) .
\eea
One can choose the coordinate system in the 4-dimensional Euclidean space-time 
in such a manner that the vectors $V_\mu $ and $W_\mu$ will have the next components
\bea
&V_\mu = ( V, 0,0,0), \,\,\,\,\, W_\mu = (W \cos \psi , W \sin \psi,0,0),
\eea
here, we introduced the angle $\psi$ between vectors $V_\mu, W_\mu$.
The knowledge of the angle dependence allows us to
restore the scalar product terms $(\X X)$ from the corresponding terms
$V W \cos \psi  $ in the final
expression after completing whole calculation. It is possible  to proceed 
through all calculations with 
the original real vectors $X_{1\mu}, X_{2\mu}$ 
as well. But for our purpose to calculate the effective Lagrangian with a constant
background the replacement defined by Eq. (28)
is more convenient. With  
these notations one can write down the one-loop contribution to the effective action as
follows
\bea
\Delta S_{eff} &=& - {\rm tr} \, \ln [ (k^2 - a )^2 ( k^2 - 2 a) + 2 \alpha ( k^2 - a ) \nn \\
      & &- ( 4 a^2 - 3 g^4 V^2W^2) ( k^2 - 2 a) \nn \\
      && -     g^2 W^2 \beta - g^2 V^2  \bar \beta] ,
\eea
where
\bea
  a &=& -g^2 q V W, \nn \\
 q &=& \cos \psi , \nn \\
 \alpha &=&  -4g^2 V W k^2cos \tone 
     (q \cos \tone + \sin \psi \sin \tone \cos \ttwo) , \nn \\
\beta &=&  -4g^2 V^2 k^2 \cos ^2 \tone ,\nn \\
 \bar \beta &=&  -4g^2 W^2 k^2 
     (q \cos \tone + \sin \psi \sin \tone \cos \ttwo)^2 .
\eea

One can calculate explicitly  the leading logarithmic part in the one-loop
 effective Lagrangian  using a cut-off regularization. After performing the 
integration over 
$k, \, \theta_1, \, \theta_2, \, \phi $ one results in 
\bea
L_{eff}  &=&-\dfrac{g^2}{4} { \chi}^2  
 -g^2 \Lambda^2 (\X_0 X_0)  \nn \\
  & & - \dfrac{11 g^4}{96 \pi^2} {\chi}^2 ( \ln [\dfrac{g^2\chi}{\Lambda^2}] 
                      -c)  + \dfrac{11 i g^4}{192 \pi} {\chi}^2 ,
\eea
where $c$ is a number parameter, and we omitted the 
quartic terms proportional to $\Lambda^4$.
To calculate the number factor  $c$ and the imaginary part of the effective Lagrangian
we used  numerical computation with
 considering  an analytical continuation in the  parameter
$q$ from $q=1$ to  $q=0$ which corresponds to analytical continuation
of the $\chi^2$ to negative values.  Since the   final 
expression possesses a covariant form one can   
calculate the number parameter $c=0.3769... $ and, what is more important, the
imaginary part of the effective Lagrangian.
After subtracting infinities within  the minimal subtraction scheme 
and proper mass and charge renormalization
we obtain a final expression for the one-loop 
renormalized effective Lagrangian with valence gluons 
\bea
{\cal L}_{eff}^{ren} 
  & =& - \dfrac{g^2}{4} { \chi}^2  
 - m_R^2 (\X_0 X_0)     \nn \\
  & & - \dfrac{11 g^4}{96 \pi^2} {\chi}^2 (\ln [\dfrac{g^2 \chi}{\mu^2}] 
           -\dfrac{1}{2}) + \dfrac{11 i g^4}{192 \pi} {\chi}^2 .
\eea

It should be stressed that the sign of the imaginary part 
is fixed by the causal structure 
which is under the control by putting the  infinithesimal quantity
 $- i \epsilon $ in propogators. For the case $q=0$ one has checked that 
the real part of the 
function 
under the logarithm in Eq. (31) is  
negative for all admissible  
values of  $k_\mu$.
This determines uniquely the sign of the imaginary part in the
effective Lagrangian.
 We can keep also the renormalised mass parameter $m_R$
since in the presence of the magnetic potential $C_\mu$ 
 the off-diagonal gluons behave
like a charged matter and the mass term does not spoil the gauge invariance.
In the case of $m_R = 0$ the vacuum energy density has
a non-trivial vacuum at the non-zero value for the classical field strength
$\chi$
\bea
<{\chi}> =\dfrac{ \mu^2}{g^2} \exp {(-\frac{24 \pi^2}{11 g^2})}.  
\eea
This minimum does not correspond to a stable vacuum
due to the presence of the imaginary part in the effective Lagrangian.

Notice, that our final expression fro the effective
Lagrangian, Eq. (34), has a manifest gauge and 
Lorentz invariant form and the number factor in front of the logarithm
leads  to a correct negative $\beta$-function of QCD in agreement
with the renormalization group equation approach.

\end{document}